# CONVERGENT BAYESIAN FORMULATIONS OF BLIND SOURCE SEPARATION AND ELECTROMAGNETIC SOURCE ESTIMATION

K. H. KNUTH and H. G. VAUGHAN, JR.
*Dynamic Brain Imaging Laboratory*
*Department of Neuroscience*
*Albert Einstein College of Medicine*
*Bronx NY 10461, USA*

**Abstract**. We consider two areas of research that have been developing in parallel over the last decade: blind source separation (BSS) and electromagnetic source estimation (ESE). BSS deals with the recovery of source signals when only mixtures of signals can be obtained from an array of detectors and the only prior knowledge consists of some information about the nature of the source signals. On the other hand, ESE utilizes knowledge of the electromagnetic forward problem to assign source signals to their respective generators, while information about the signals themselves is typically ignored.

We demonstrate that these two techniques can be derived from the same starting point using the Bayesian formalism. This suggests a means by which new algorithms can be developed that utilize as much relevant information as possible. We also briefly mention some preliminary work that supports the value of integrating information used by these two techniques and review the kinds of information that may be useful in addressing the ESE problem.

## 1. Introduction

Determination of the spatial and temporal behavior of neural activity using non-invasive methods is critical to the study of human brain function. While there are several experimental techniques that provide information on neural activity, no single method currently provides information with the spatial and temporal resolution necessary to delineate the neural processes that underlie human cognition and sensorimotor function. Electroencephalography (EEG) and magnetoencephalography (MEG) provide sub-millisecond temporal resolution of neuroelectric current flow. However, the electric potentials and magnetic fields recorded by the detectors on the surface of the head consist of a linear superposition of the signals emitted by the active neural sources. It is well known that in the absence of additional constraints, solutions to the electromagnetic inverse problem, which consist of the determination of the underlying neural current distributions given the surface potentials and fields, are not unique. The problem increases in difficulty as the number of neural sources increases. The limited ability of current methods to accurately and consistently separate and localize the neuro-electromagnetic sources has impaired confidence in the value of surface-recorded electromagnetic data for examining the dynamics of the underlying neural activity that generates them. We propose an approach to the electromagnetic inverse problem that combines information that is typically utilized in each of two methods that have been developed in parallel over the past decade: namely, blind source separation (BSS) and electromagnetic source estimation (ESE). The additional prior information that can be brought to bear through a composite approach should facilitate the development of algorithms that can achieve more accurate definition of the intracranial signal sources and the source waveforms.

Much progress has been made in the last decade to understand and attain the segregation of mixed signals using blind source separation (BSS) techniques. The success of the application of BSS toward specific problems is often surprising given the apparently minimal amount of information used to achieve separation. This minimal information usually consists of the assumptions of independence of the source signals and that the source amplitude densities can be described by a particular class of probability distributions.

Electromagnetic Source Estimation (ESE) attempts to locate and characterize the neural signals responsible for the electric potentials and magnetic fields recorded on the surface of the head during synchronized neural activity. Most ESE techniques utilize information regarding the electric properties of typical neural generators, the locations of the detectors, and the properties of propagation of electric currents and magnetic fields from the sources to the detectors.

While some anatomical and physiological information may be incorporated in ESE algorithms, typically very little information about the signals themselves is included. Source signal solutions are constrained to be decorrelated or statistically independent, but no additional information about the source amplitude densities, typical amplitudes of the various sources, and stochastic variations in the source waveforms is included.

## 2. The Bayesian Approach

The general problems of source separation and source localization are inherently inference problems because there is not sufficient information to deduce a solution. Bayesian inference allows one to calculate the probability that a given hypothesis, or model, is correct given the data and any prior information. We demonstrate that the solutions to both of these problems can be derived from the same basic starting point, Bayes' Theorem. These derivations will diverge when we focus on particular parameters to model the physical situation, incorporate additional information, and make simplifying assumptions.

Given a model of a physical problem of interest and some additional data, we can use Bayes' Theorem to describe how this additional information affects our belief as to whether a particular model provides an accurate description:

$$P(model \mid data, I) = \frac{P(data \mid model, I) \, P(model \mid I)}{P(data \mid I)} \qquad (1)$$

where *I* represents any prior information we may have.

In the case of the neural source estimation problem, there are several parameters that can be used to describe the model: the source positions **p** and orientations **q**, the source signals **s**(t), and the mixing matrix **A**, which describes how the source signals are mixed when recorded by the detectors. The data consist of the electromagnetic signals recorded by our detectors. Representing the recorded mixed signals as **x**(t), we rewrite Equation (1) as a proportionality

$$P(\mathbf{p}, \mathbf{q}, \mathbf{A}, \mathbf{s}(t) \mid \mathbf{x}(t), I) \propto P(\mathbf{x}(t) \mid \mathbf{p}, \mathbf{q}, \mathbf{A}, \mathbf{s}(t), I) P(\mathbf{p}, \mathbf{q}, \mathbf{A}, \mathbf{s}(t) \mid I). \qquad (2)$$

The prior probability on the right-hand side can be factored into several terms

$$P(\mathbf{p}, \mathbf{q}, \mathbf{A}, \mathbf{s}(t) \mid \mathbf{x}(t), I) \propto$$
$$P(\mathbf{x}(t) \mid \mathbf{A}, \mathbf{s}(t), I) P(\mathbf{A} \mid \mathbf{p}, \mathbf{q}, I) P(\mathbf{q} \mid \mathbf{p}, I) P(\mathbf{p} \mid I) P(\mathbf{s}(t) \mid I). \qquad (3)$$

The first term on the right, the likelihood, has been simplified since the likelihood of the data **x**(t) depends explicitly on the matrix **A** and the source signals **s**(t) such that the mixing is linear and instantaneous, $\mathbf{x}(t) = \mathbf{A}\,\mathbf{s}(t)$. The second term, the prior $P(\mathbf{A} \mid \mathbf{p}, \mathbf{q}, I)$, describes the dependency of the mixing matrix on the positions and orientations of the sources. The second prior $P(\mathbf{q} \mid \mathbf{p}, I)$ describes the probability of the orientations of the sources given their positions. Prior knowledge regarding possible source positions is expressed in the third prior $P(\mathbf{p} \mid I)$. Since the signal propagation does not depend on the source amplitude and the source behaviors are at present not known to depend on source position, the source signal prior $P(\mathbf{s}(t) \mid I)$ depends only on our prior expectations of the nature of the signals themselves.

### 2.1. Bayesian Formulation of BSS

BSS endeavors to separate the source signals without recourse to information on their spatial distribution and relation to the detectors. Two basic approaches are used in blind source separation: algebraic approaches, which are higher-order generalizations of singular value decomposition (SVD) or principal components analysis (PCA); and information-theoretic approaches, which utilize maximum entropy, maximum likelihood, or minimum mutual information.

To illustrate the application of the Bayesian formulation to the latter class of approaches, we focus on a popular technique, the Bell-Sejnowski Independent Component Analysis (ICA) algorithm (Bell & Sejnowski 1995). The algorithm is cast in the form of a neural network that takes the mixed signals, performs a nonlinear transformation on them and then separates them by maximizing the entropy of the output signals so that they are statistically independent. Viewed differently, this algorithm uses information about the probability distribution of source amplitudes (via the nonlinear transformation) and the statistical independence of the source signals to separate the signals.

Derivation of the Bell-Sejnowski ICA algorithm using the Bayesian methodology has been described elsewhere (Knuth 1998a,b) and after the initial assumptions are made, the remainder of the derivation is similar to those presented by MacKay (1996) and Pearlmutter and Parra (1996). We will outline the derivation, paying special attention to the particular assumptions that distinguish the results from ESE.

BSS assumes that we know nothing about the sources or the nature of the mixing. This lack of information can be expressed by assigning uniform distributions to the priors, $P(\mathbf{A} \mid \mathbf{p}, \mathbf{q}, I)$, $P(\mathbf{q} \mid \mathbf{p}, I)$ and $P(\mathbf{p} \mid I)$ that extend over a large enough range to account for all possible values of the model parameters. We then marginalize over the model parameters $\mathbf{p}$ and $\mathbf{q}$ to obtain

$$P(\mathbf{A}, \mathbf{s}(t) \mid \mathbf{x}(t), I) \propto P(\mathbf{x}(t) \mid \mathbf{A}, \mathbf{s}(t), I) P(\mathbf{s}(t) \mid I), \qquad (4)$$

where all constants are absorbed by the proportionality.

If we believe the external noise level to be sufficiently small then the inverse of the mixing matrix $\mathbf{A}$ can be used to obtain the source signals. The source signals $\mathbf{s}(t)$ can be treated as nuisance parameters and removed by marginalizing over $\mathbf{s}(t)$

$$P(\mathbf{A} \mid \mathbf{x}(t), I) \propto \int d\mathbf{s} \quad P(\mathbf{x}(t) \mid \mathbf{A}, \mathbf{s}(t), I) P(\mathbf{s}(t) \mid I). \qquad (5)$$

To reflect the statistical independence of the sources, the prior $P(\mathbf{s}(t) \mid I)$ is written as a product of prior probabilities for each source. Typically a hyper-Gaussian probability density, such as the derivative of a sigmoid function, is assigned to describe the source amplitude density of the source signals. Of course, if more is known about the dynamical behavior of the source signals, one can assign more detailed priors. Finally, if we assume that the mixing is linear, instantaneous and relatively noise-free, as described above, delta functions can be assigned to the likelihood. This results in

$$P(\mathbf{A} \mid \mathbf{x}(t), I) \propto \int d\mathbf{s} \quad \prod_i \delta(x_i - A_{ik} s_k) \prod_l p_l(s_l) \qquad (6)$$

where the Einstein summation convention is used to denote matrix multiplication. The delta functions allow one to evaluate the multi-dimensional integral. The most probable value for the matrix $\mathbf{A}$, is found by maximizing the logarithm of the probability

$$\log P(\mathbf{A} \mid \mathbf{x}(t), I) = -\log \det \mathbf{A} + \sum_l p_l\left(A_{lk}^{-1} x_k\right) + C. \qquad (7)$$

The Bell-Sejnowski ICA algorithm is constructed by using a stochastic gradient search to estimate the separation matrix, $\mathbf{W} = \mathbf{A}^{-1}$, which maximizes Equation (7). This algorithm has been shown to work very well in separating mixed signals where the prior source amplitude densities provide an adequate representation of our knowledge about the source behavior.

### 2.2. Bayesian Formulation of ESE

In this section we proceed with a derivation of one of the most common algorithms used in ESE. This algorithm takes a very different approach to solving the source separation problem by utilizing information ignored by ICA and ignoring information used by ICA. We focus on the model parameters describing the positions and orientations of the sources as well as the source signals. Additional information takes the form of knowledge regarding the propagation of the signals from the sources to the detectors. This consists of a description of the forward model, expressed as a function, $A_{ij} = F(d_i, p_j, q_j)$ determined either analytically or numerically, that determines the matrix elements of **A** given the source and detector positions and orientations, and a model of the volume conductor.

We begin with Equation (3), and assign a delta function to the prior probability $P(\mathbf{A} \mid \mathbf{p}, \mathbf{q}, I)$,

$$P(\mathbf{p}, \mathbf{q}, \mathbf{A}, \mathbf{s}(t) \mid \mathbf{x}(t), I) \propto$$
$$P(\mathbf{x}(t) \mid \mathbf{A}, \mathbf{s}(t), I) P(\mathbf{q} \mid \mathbf{p}, I) P(\mathbf{p} \mid I) P(\mathbf{s}(t) \mid I) \prod_{i,j} \delta(A_{ij} - F(d_i, p_j, q_j)). \quad (8)$$

Treating the matrix **A** as a nuisance parameter we marginalize over **A**,

$$P(\mathbf{p}, \mathbf{q}, \mathbf{s}(t) \mid \mathbf{x}(t), I) \propto$$
$$P(\mathbf{q} \mid \mathbf{p}, I) P(\mathbf{p} \mid I) P(\mathbf{s}(t) \mid I) \int d\mathbf{A} \ P(\mathbf{x}(t) \mid \mathbf{A}, \mathbf{s}(t), I) \prod_{i,j} \delta(A_{ij} - F(d_i, p_j, q_j)). \quad (9)$$

Unlike the previous derivation, we assign a Gaussian probability density to the likelihood function. This Gaussian density expresses that fact that we do not believe the experimental data to be precisely described by the model, and that we know the value of the expected squared deviation between the predicted and experimentally obtained data. Focusing on the integral on the right-hand side of Equation (9), we write

$$\int d\mathbf{A} \ \prod_{k,t} Exp\left[-\frac{(x_{kt} - A_{kl} s_{lt})^2}{2\sigma_k^2}\right] \prod_{i,j} \delta(A_{ij} - F(d_i, p_j, q_j)), \quad (10)$$

where we explicitly express the vector of time series data $\mathbf{x}(t)$ as $x_{kt}$ where k denotes the $k^{th}$ data channel and t denotes the $t^{th}$ time point. Evaluating the integral results in

$$P(\mathbf{p}, \mathbf{q}, \mathbf{s}(t) \mid \mathbf{x}(t), I) \propto P(\mathbf{q} \mid \mathbf{p}, I) P(\mathbf{p} \mid I) P(\mathbf{s}(t) \mid I) \prod_{k,t} Exp\left[-\frac{(x_{kt} - F_{kl} s_{lt})^2}{2\sigma_k^2}\right]. \quad (11)$$

where $F_{kl}$ represents the function $F(d_k, p_l, q_l)$. Notation can be further simplified by introducing, $\hat{x}_{kt} = F_{kl} s_{lt}$.

The three remaining priors are all very important in that they describe our prior knowledge regarding the positions and orientations of the sources and details of the source signals. The first two priors, if not entirely neglected, are assigned functions that constrain solutions to lie on a cortical surface with a particular orientation. The prior that represents information about source signals is typically neglected. In this derivation, we will assume complete ignorance by assigning a uniform probability distribution to each of the three priors. These uniform distributions take the form of constants, which are absorbed by the proportionality. Since the ultimate goal is to find the most probable solution, we look for the parameters that maximize the logarithm of the probability

$$\log P(\mathbf{p}, \mathbf{q}, \mathbf{s}(t) | \mathbf{x}(t), I) = -\sum_k \sum_t \frac{(x_{kt} - \hat{x}_{kt})^2}{2\sigma_k^2} + C, \quad (12)$$

which is equivalent to minimizing the "cost function"

$$\chi^2 = \sum_k \sum_t \frac{(x_{kt} - \hat{x}_{kt})^2}{\sigma_k^2}. \quad (13)$$

This is the chi-squared cost function that is typically used to evaluate source models in ESE. When the variances associated with the different detector channels are identical, the expression above reduces to a minimization of the least-squared error between the data and their predicted values. It is important to remember that this result is derived from the assumptions of linear instantaneous mixing, ignorance of possible source locations and orientations, statistical independence of the source activity, and a Gaussian likelihood that the data are predicted by the model.

3. **Incorporation of prior information in ESE**

It is important to compare the derivations of the solutions to the problems of BSS and ESE and pay special attention to the information that is used through the assignment of the relevant prior probabilities. In most cases, the researcher has several different kinds of additional information or prior knowledge about the specific physical problem that can be naturally integrated using the Bayesian approach. In a previous paper, we derived an algorithm similar to ICA that includes information about the signal propagation, the original source positions and the signal properties (Knuth 1998b). Specifically, we examined an artificial problem where the signal amplitudes have a one over distance squared fall-off, and the prior source positions are described by three-dimensional Gaussian probability densities. The algorithm was shown to separate signals that are difficult to separate using ICA in conjunction with a hyper-Gaussian source density. While this preliminary work demonstrated that separation is possible using additional accurate information in conjunction with inaccurate information regarding source behavior, we stress that the information utilized should be as accurate as possible. In addition, some types of information may be more useful than others.

Studies must be performed to understand the relative importance of the different types of information.

We now discuss several sources of information of specific relevance to neuro-electrophysiologic source separation. It is convenient to divide the prior information into two classes of variables, namely those associated with the source model and those related to the volume-conductive head model. The former includes information on the time varying electromagnetic signals themselves, as well as information on the number, geometry and spatial location of the signal sources. The latter includes information on the shape of the brain and its covering and the conductivities of the various tissues.

We first consider information about the source signals themselves, which are ordinarily represented by time-varying current dipole approximations of the complex patterns of synchronous neural activity within a macroscopic source. The dipolar nature of the source model allows constraints on directionality imposed by the orientation of the dipole, and on maximum amplitudes and temporal amplitude sequences of the dipole moments as well. Due to their small amplitude relative to the ongoing spontaneous brain activity, the signals of interest, so-called "event-related" potentials or fields, are enhanced relative to the background activity by signal averaging. These event-related signals are typically transients that are time-locked to an external event, usually a stimulus or motor response. Although the residual background activity is generally called "noise", its spatial and temporal properties have a statistical structure that can be empirically determined. As yet, little attention has been paid to characterizing the statistics of the residual spontaneous activity in the event-related averaged signals of interest. This residual "noise" is, however, a major source of error in ESE, and needs to be dealt with in a more direct fashion.

As previously noted, ESE approaches have rarely included information about the signals themselves. In another paper (Knuth 1998a), we discussed the inability of ICA to separate EEG/MEG signals due to the fact that the hyper-Gaussian densities used by the algorithm do not accurately represent EEG/MEG amplitude densities. These are typically multi-modal and thus require many parameters to model. We require prior probability densities that more accurately describe what is known about the source waveforms. These priors can also describe dynamical information through the assignment of prior probabilities conditional on the derivatives of the signals.

Information regarding the possible positions and orientations of neural sources can be obtained with millimeter resolution from MR images. It is known that the equivalent dipoles that represent the net extracellular current flow within a cortical element are normal to the cortical surface due to the cellular architecture of the cortex. This provides a strong constraint on both cortical source locations and orientations (Hamalainen and Ilmoniemi 1984; George et al. 1991; Dale and Sereno 1993). The Bayesian formulation is particularly advantageous in providing a means by which this information can be included as accurately as possible. Thus, the cortical surface can be assigned rather precise priors that reflect our knowledge of the cortical geometry, whereas deeper, less-understood structures can be assigned uniform priors that reflect our ignorance. In this way, the entire brain is treated in a uniform manner with different probabilities reflecting knowledge specific to each region.

Determination of the number of neural sources, also known as model order, has been an extremely difficult problem in source estimation. While the number of sources is typically estimated from the number of significant singular values of the SVD of the data matrix or its covariance matrix, it is known that this provides a minimum estimate of the model order. The Bayesian approach provides a natural means by which one can perform model order comparisons. By explicitly calculating the probabilities of each model, the effect of changes in model order can be quantitatively assessed.

Turning to the head model variables, the finite conductive volume of the head constrains the equations describing the potentials and fields detected by surface recordings (i.e. the forward problem). Typically used are head models in which the brain, extra-cerebral fluid, skull and scalp are modeled as concentric spheres. These simplified volume conductor models, coupled with the current dipole model of the neural sources, permit analytic solutions that describe the potentials and fields on the surface of the head. More realistic head models take into account individual brain and skull geometry as indicated by magnetic resonance images (MRI). In these models, numerical techniques are required to estimate the surface effects of a neural source. Conductivities of the different tissues are estimated from sparse empirical data and thus are not accurately known.

While the importance of incorporating all possible sources of information to aid in solving the neuro-electromagnetic inverse problem is well appreciated, the methodology is ordinarily limited to second-order statistics. An excellent work by Dale and Sereno (1993) outlines in detail the incorporation of EEG, MEG and MRI using second-order techniques. When Gaussian priors accurately describe our knowledge, these techniques are optimal. However, when this is not the case these second-order techniques are unduly restrictive.

We comment on how these ideas can be better understood and improved when examined from the Bayesian viewpoint. One of the most common misconceptions is that the Gaussian likelihood implies the existence of a physical signal consisting of independent, additive Gaussian noise. Recall that the assignment of the likelihood represents what is known about the predictive capability of a model. The assignment of the Gaussian likelihood in Equation (10) is appropriate when our prior information consists of the value of the expected squared deviation between the predicted and experimentally observed data values. In short, the variance in the Gaussian density represents our uncertainty that the model predicts the data. This uncertainty could be due to instrument noise, signals from additional or unexpected neural sources, or simplifying approximations in the description of the physical situation. Typically in the case of stimulus-elicited neuro-electromagnetic signals, instrument noise and neural background activity are estimated from the prestimulus baseline signals. While instrument noise may not exhibit correlations across channels, neural background activity often does. Detailed studies regarding the nature of the neural background will assist in a better description of the uncertainties due to these signals. More importantly, model mis-specification contributes uncertainties to the likelihood via the assignment of the delta function prior, describing a naive confidence in the approximate transfer functions (see also Dale and Sereno 1993). Research performed on the effects of model mis-specification (Cuffin 1993; Zhang and Jewett 1993) can aid in the assignment of a more accurate mixing matrix prior, which better describes the known limitations of our approach.

It is well known that EEG and MEG do not provide identical information about neuroelectric source currents, even though they reflect the same neural events. MEG is thought to represent primarily intracellular currents, whereas EEG is generated by extracellular volume currents. MEG is also insensitive to currents generated in radial dipolar sources, whereas EEG represents sources in all orientations. Thus, each method has its significant strengths and weakness as indices of intracranial neural activty. EEG and MEG data can be combined by augmenting the EEG mixing matrix with the MEG mixing matrix. Known differences in instrument noise and background activity recorded using these two techniques can be represented by differences in their corresponding likelihoods. In addition, information about model mis-specification for the two techniques can be accounted for in their respective mixing matrix priors.

Integration of information obtained from functional imaging, which is based on metabolic effects due to the activity of neural tissue, with ESE has received considerable recent attention. Due to the presumably better spatial accuracy of metabolic imaging methods, it has been suggested that electromagnetic sources might be constrained in location by fMRI or PET activation. These suggestions must be viewed with considerable caution inasmuch as the metabolic effects, which are visualized as differences between two states of brain activation, are by no means directly comparable to the electromagnetic indices of neural activity. Not only is the time scale vastly different – seconds versus milliseconds, but the quantitative coupling between the measures has not yet been defined. Metabolic imaging and electromagnetic source imaging techniques may index sufficiently different physiological processes, so that information obtained using one technique may not provide useful constraints on solutions based on the other technique.

In conclusion, we have laid out a Bayesian framework within which the ESE problem may be profitably explored, and provides both the means of incorporating a broad range of information with varying degrees of confidence, and for testing the probability of various alternative solutions to this inverse problem.

## ACKNOWLEDGEMENTS

This work was supported by NIH 5P50 DC00223, NIH 5P30 HD01799, and NIH NIDCD 5 T32 DC00039-05